\documentclass[%
 reprint,
superscriptaddress,
 amsmath,amssymb,
 aps,
 pra,
]{revtex4-1}

\usepackage{siunitx}
\usepackage{soul}
\usepackage[T1]{fontenc}
\usepackage[latin2]{inputenc} 
\usepackage{floatrow}
\usepackage{graphicx}
\usepackage{dcolumn}
\usepackage{bm}
\usepackage[bookmarksopen,bookmarksdepth=3,colorlinks=true,linkcolor=blue,citecolor=blue,urlcolor=blue]{hyperref}

\begin{document}


\title{Reduced thermal conductivity of TiNiSn/HfNiSn superlattices}

\author{Paulina Ho\l{}uj}
 \email{holuj@uni-mainz.de}
 \affiliation{Institute of Physics, University of Mainz, Staudinger Weg 7, 55128 Mainz, Germany.}
 \affiliation{Graduate School Materials Science in Mainz, Staudinger Weg 9, 55128 Mainz, Germany.}

\author{Christoph Euler}
 \affiliation{Institute of Physics, University of Mainz, Staudinger Weg 7, 55128 Mainz, Germany.}

\author{Benjamin Balke}
 \affiliation{Institute of Inorganic and Analytical Chemistry, University of Mainz, Duesbergweg 10-14, 55128 Mainz, Germany.}

\author{Ute Kolb}
 \affiliation{Institute for Physical Chemistry, University of Mainz, Welderweg 11, 55099 Mainz, Germany.}

\author{Gregor Fiedler}
 \affiliation{Faculty of Physics, University of Duisburg-Essen, Lotharstra{\ss}e 1, 47048 Duisburg, Germany.}

\author{Mathis M. M\"uller}
 \affiliation{Institute of Applied Geosciences, TU Darmstadt, Schnittspahnstra{\ss}e 9, 64287 Darmstadt, Germany.}

\author{Tino Jaeger}
 \affiliation{Institute of Physics, University of Mainz, Staudinger Weg 7, 55128 Mainz, Germany.}

\author{Peter Kratzer}
 \affiliation{Faculty of Physics, University of Duisburg-Essen, Lotharstra{\ss}e 1, 47048 Duisburg, Germany.}

\author{Gerhard Jakob}
 \affiliation{Institute of Physics, University of Mainz, Staudinger Weg 7, 55128 Mainz, Germany.}
 \affiliation{Graduate School Materials Science in Mainz, Staudinger Weg 9, 55128 Mainz, Germany.}

\date{\today}

\begin{abstract}
Diminution of the thermal conductivity is a crucial aspect in thermoelectric research. We report a systematic and significant reduction of the cross-plane thermal conductivity in a model system consisting of DC sputtered TiNiSn and HfNiSn half-Heusler superlattices. The reduction of $\kappa$ is measured by the 3$\omega$ method and originates from phonon scattering at the internal interfaces. Heat transport in the superlattices is calculated based on Boltzmann transport theory, including a diffusive mismatch model for the phonons at the internal interfaces. Down to superlattice periodicity of 3 nm the phonon spectrum mismatch between the superlattice components quantitatively explains the reduction of $\kappa$. For very thin individual layers the interface model breaks down and the artificial crystal shows an enhanced $\kappa$. We also present an enhanced \textit{ZT} value for all investigated superlattices compared to the single TiNiSn and HfNiSn films. 

\end{abstract}

\keywords{thermoelectrics, half-Heusler, 3\omega method, thermal conductivity, superlattice}
\maketitle


The ability to convert a temperature difference to electricity as well as the possibility of both heating and cooling are very valuable properties of thermoelectric materials. Despite these promising features they are not widely used in industry as they posses low efficiency, described by a dimensionless figure of merit $ZT=\frac{S^{2}\sigma}{\kappa_{\mathrm{tot}}}T$ ($S$ - Seebeck coefficient, $\sigma$ - electrical conductivity, $\kappa_{\mathrm{tot}}$ - thermal conductivity, \textit{T} - absolute temperature).

\textit{M}NiSn half-Heusler (HH) materials, where \textit{M} = (Ti, Hf, Zr) are considered to be materials having high potential for thermoelectric (TE) applications due to Seebeck coefficients in the range of $-200\ \mu$V/K for bulk materials already at room temperature \cite{Uher99}. However, simultaneously they exhibit thermal conductivities even as high as 10 W/(mK) preventing the achievement of a satisfactory value of \textit{ZT}, which is a measure of the material's applicability \cite{Uher99}. Nowadays the scientific community is exploring all means to reduce the thermal conductivity of HH materials in order to enhance their \textit{ZT}. Promising approaches include: the introduction of grain boundaries by melt spinning or ball milling processes and further spark plasma sintering \cite{Gelbstein11}, phase separation during the solidification of bulk materials \cite{Rausch14, Kimura09, Populoh12, Schwall13}, or the thin film and superlattice (SL) approach \cite{Jaeger14, Wang10}. The latter was investigated by Venkatasubramanian et al.\ for the Bi$_{2}$Te$_{3}$/Sb$_{2}$Te$_{3}$ SL system and resulted in the highest ever reported \textit{ZT} value of 2.4 at room temperature \cite{Venkatasubramanian00, Venkatasubramanian01}. In contrast to tellurides, HH materials posses the peak of efficiency at high temperature (above 700 K) \cite{Graf11}. Moreover, the cost of bulk HHs is about one order of magnitude lower compared to the former compounds \cite{LeBlanc14}.

The superlattice approach is a promising way to reduce the thermal conductivity, however, the total thermal conductivity consists of an electronic and a phononic contribution ($\kappa_{\mathrm{tot}}=\kappa_{\mathrm{el}}+\kappa_{\mathrm{ph}}$). The first is directly related to the electrical conductivity via the Wiedemann-Franz law ($\frac{\kappa_{\mathrm{el}}}{\sigma T}=const$) \cite{Kittel}. Therefore, reduction of $\kappa_{\mathrm{el}}$ leads to a reduction of $\sigma$ and to a decrease of \textit{ZT}. To ensure that $\kappa_{\mathrm{el}}$ and other electronic properties remain unchanged, isoelectronic elements are used as substitutes for the \textit{M} element in the compounds so that mainly the phononic thermal conductivity is affected. In our model system SLs have constant total thicknesses, but variable SL period and therefore number of interfaces. However, SLs with ultrashort period consist only of a few atomic planes of the different materials. Thus they should better be considered not as layers of material 1 and material 2 separated by interfaces, but rather as an artificial tailor made new material with a large crystallographic unit cell.

\begin{figure*}
 \includegraphics[width=6.75in]{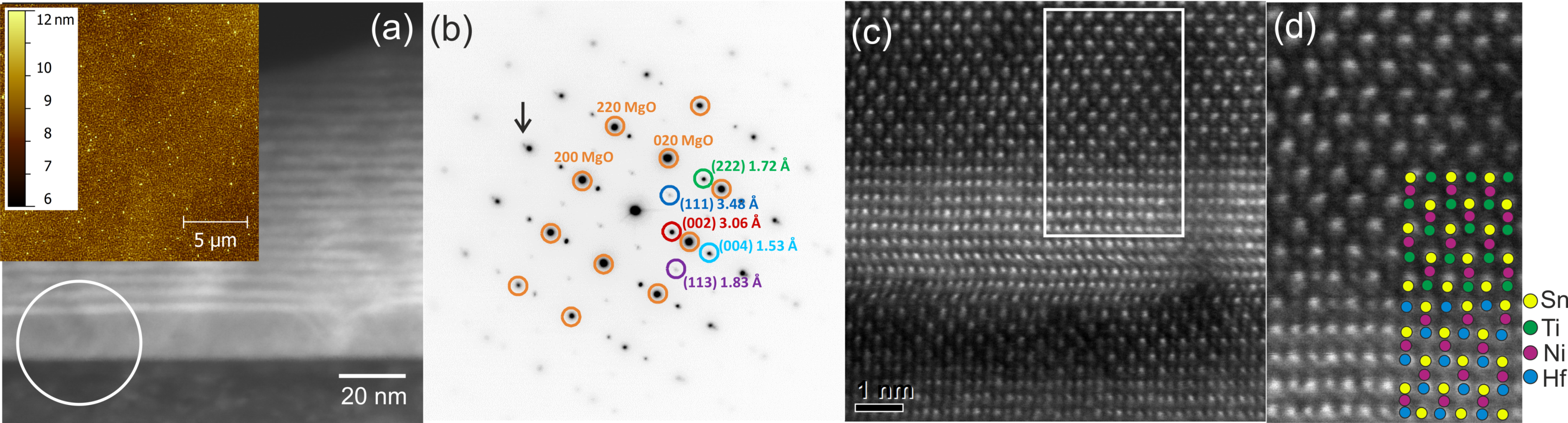}
 \caption{(a) STEM dark field image of a SL cross-section with a period of 4 nm, composed of TiNiSn (dark contrast) and HfNiSn (bright contrast) and surface topography of an approximately 800 nm thick SL with rms roughness of 1.1 nm (AFM image) in inset, (b) SAED of a circular region from the image (a), (c) high-resolution STEM image, (d) the magnification of a rectangular region with the assignment of atoms.}
 \label{fig:TEM}
\end{figure*}

TiNiSn/HfNiSn SLs having the same amount of both materials in one period were grown by DC magnetron sputtering. Details about sample preparation are presented in the supplementary data. The cross-plane thermal conductivity of HH SLs was measured with the 3$\omega$ method \cite{Cahill90, Cahill02, Borca-Tasciuc01}. The sample geometry and the measurement procedure was described in detail in our previous work \cite{Jaeger14} and in the supplementary materials.

A crucial aspect of the approach pursued here is a strong sensitivity to the quality of the interfaces.  \hbox{X-ray} diffraction (XRD), atomic force microscopy (AFM) and scanning transmission electron microscopy (STEM) were used to estimate interface quality. Assuming that the interface roughness transfers to the surface roughness, a root mean square (rms) roughness of 1.1\,\,nm for a 800\,\,nm thick SL (Fig.~\ref{fig:TEM}a inset) implies a smooth boundary between layers. Moreover, the STEM image shown in Fig.~\ref{fig:TEM}a proves the ability to grow high quality structures. In this particular case, a SL with a period of about 4\,\,nm is presented. Every bright layer of the stack corresponds to HfNiSn, whereas dark layers demonstrate the presence of TiNiSn. 

Additionally, selected area electron diffraction (SAED) of a circular region marked in Fig.~\ref{fig:TEM}a was measured and the result, together with the assignment of crystalline directions to the reflections, is presented in Fig.~\ref{fig:TEM}b. Every orange circle indicates a MgO diffraction spot, whereas other colors refer to film reflections. Observed diffraction spots confirm strongly directional growth of the film according to the orientation provided by the substrate. Due to the slight difference between lattice constants of both HH materials, we also observe a splitting of the reflection indicated by an arrow. Unequal lattice constants lead to the distortion of the lattice visible at the high-resolution STEM image (Fig.~\ref{fig:TEM}c). Based on the brightness contrast, one can clearly recognize piles of the constituent atoms.

\begin{figure}
 \includegraphics[width=3.375in]{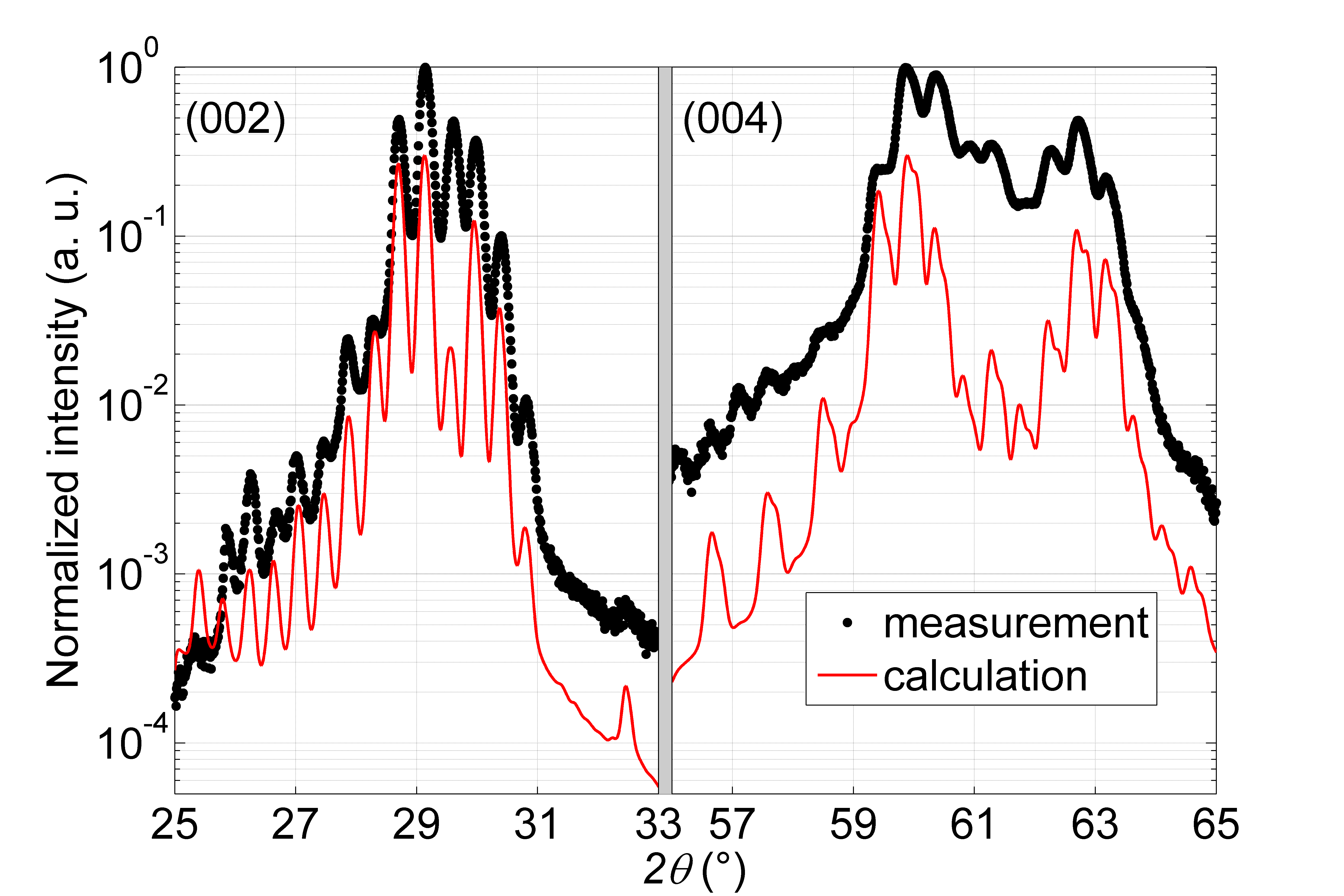}
 \caption{Measured and calculated XRD $\theta-2\theta$ profile of a SL with period 21.9 nm. The intensities are rescaled to create an offset for clarity. Shown are the regions around (002) and (004) reflection.}
 \label{fig:XRD}
\end{figure}

While AFM and STEM yield local information, XRD averages over a large area and is sensitive to inner interfaces, since the films are thinner than the penetration depth. Typical XRD data is shown in Fig.~\ref{fig:XRD}. The [001] ([100]) direction of the film is parallel to the [001] ([110]) direction of the MgO substrate as is evident by the occurrence of (002) and (004) film peaks. Additional satellites around the main diffraction lines of \hbox{TiNiSn} and HfNiSn arise from the extra periodicity present in the samples \cite{Fullerton92}. These satellite peaks are particularly sensitive to the quality of the boundaries between the constituent layers as any continuous fluctuations would cause broadening of all SL peaks resulting in difficulties to resolve additional oscillations \cite{Fullerton92}. Consequently, pronounced satellites imply better sample quality. 

\begin{figure*}
\floatbox[{\capbeside\thisfloatsetup{capbesideposition={right,top},capbesidewidth=4cm}}]{figure}[\FBwidth]
 {\includegraphics[width=5in]{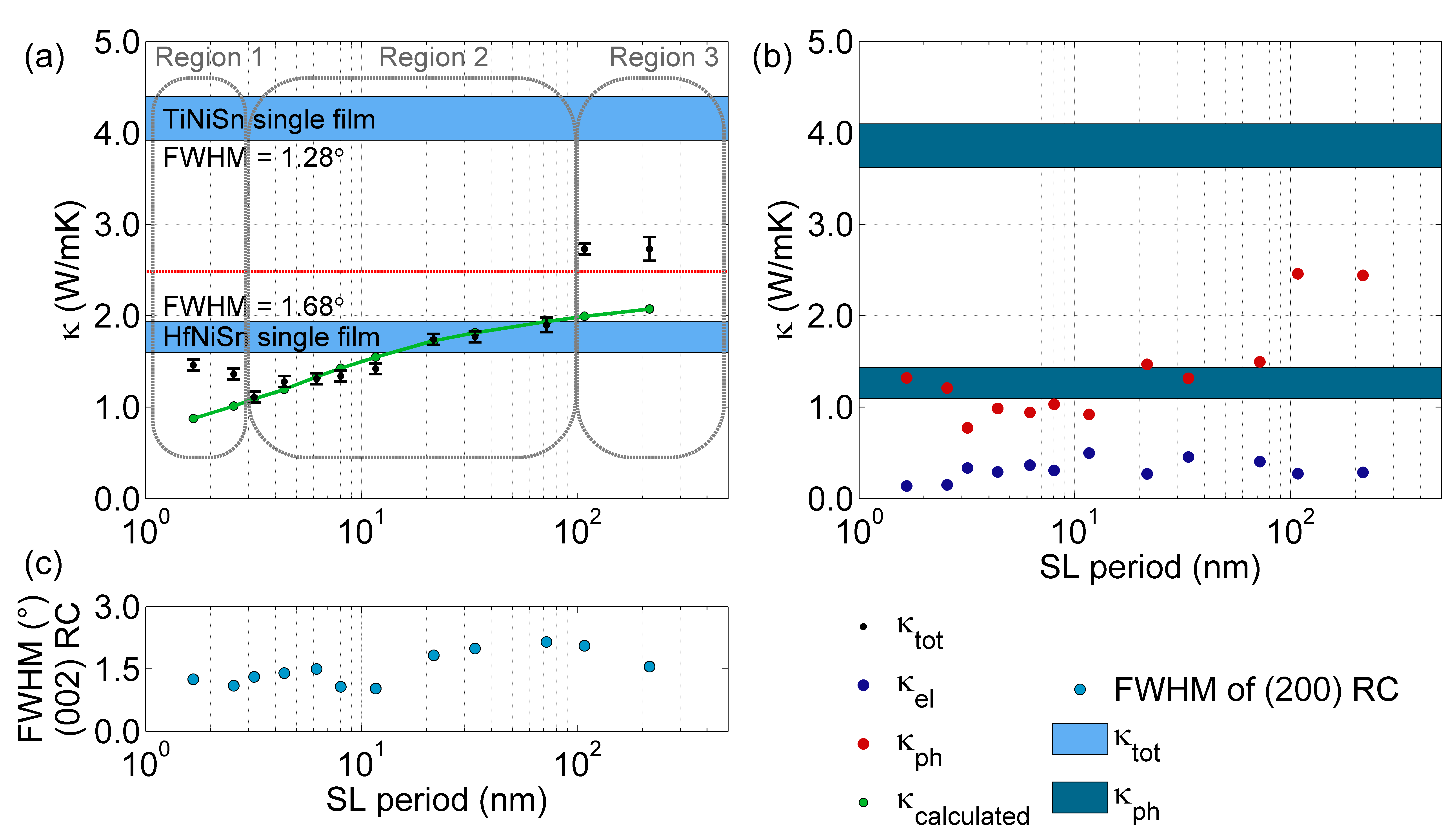}}
 {\caption{(a) Measured cross-plane thermal conductivity of HH SLs with varying period (black dots) and $\kappa_{\mathrm{calculated}}$ based on a series of thermal resistances (green dots). Experiment and calculation agree for SL periods in region 2. (b) Electronic and phononic part of the thermal conductivity for every SL and single films. (c) FWHM of the (002) rocking curves of the films and SLs.} \label{fig:3w}}
\end{figure*}

In Fig.~\ref{fig:XRD}, besides measured XRD data, the calculation based on coherent x-ray scattering for a perfect SL is presented. To give a realistic physical shape to the simulated spectra, peaks were convoluted with the measured profiles of epitaxial TiNiSn films. Convolution parameters resulted from the fit of two Pseudo-Voigt profiles to (002) and (004) peaks for single \hbox{TiNiSn} films, respectively. Scattering factors of the atoms in the planes were taken from literature \cite{CrystTables} and lattice plane spacing was adjusted to account for epitaxial strain. Therefore, instead of bulk lattice parameters of HH films ($a_\mathrm{{TiNiSn}} = 5.941$ \si{\angstrom}, $a_\mathrm{{HfNiSn}} = 6.083$ \si{\angstrom}) \cite{Jeitschko70}, the out-of-plane lattice parameters taken for the calculation were equal to $a_\mathrm{{TiNiSn}} = 5.906$ \si{\angstrom} and $a_\mathrm{{HfNiSn}} = 6.184$ \si{\angstrom}. The thicknesses of TiNiSn and HfNiSn were set to 10.6 nm and 11.3 nm, respectively. Comparing measured and calculated spectra of the SLs, it can be concluded that not only the position of the peaks, but also the overall peak shape agree well. However, relative intensities of the peaks are not exactly reproduced, especially for higher order satellites, indicating disturbances from the perfect periodicity in the experimental samples. 

To measure the cross-plane thermal conductivity the differential 3$\omega$ method was used. The $U_{3\omega}$ in the heater structure was measured by a SR850 DSP lock-in amplifier from Stanford Research Systems. For every SL period two sets of regular and reference samples were prepared in order to confirm the reproducibility. Based on both pairs of results the error bars of data points were estimated. The difference between the thermal conductivities of both series of samples having the same SL period was compared with an uncertainty originating from the measurement setup. Then the error bars were assumed to be equal to the greater of these values.

Besides the measurement of the thermal conductivity, the XRD rocking curves (RC) of the (002) diffraction peak or the most intense satellite were acquired for one thick sample of every studied SL period. RCs were fitted with a Pseudo-Voigt profile and their full width at half maximum (FWHM) for every period was summarized in Fig.~\ref{fig:3w}c. In contrary to our previous study on (Zr,Hf)NiSn/TiNiSn superlattices \cite{Jaeger14}, we do not see any correlation between the width of RCs and the thermal conductivity in the TiNiSn/HfNiSn system which indicates the improved growth of the sytem investigated here. 

The study of room-temperature thermal conductivity versus the SL period is presented in Fig.~\ref{fig:3w}a. The horizontal blue bars indicate the thermal conductivity of the individual film materials shown for comparison. For \hbox{TiNiSn} films we obtained an average of 4.16 W/(mK). Due to lower roughness and improved sample crystalline alignment in the current sample series this value is almost 50\% higher than the one reported by us earlier (2.8\,W/(mK)) \cite{Jaeger14}. The current result is also much closer to the bulk value of the thermal conductivity of TiNiSn reported in the literature, which ranges between $(4.6 - 9.3)\,\textrm W /(\textrm {mK})$ for arc melted \cite{Birkel12, Gelbstein11, Hohl99}, 4.8\,\,W/(mK) for microwave prepared \cite{Birkel12} and 7.5\,W/(mK) for levitation melted materials \cite{Douglas12}. Conversely, we report a reduced thermal conductivity down to 1.77\,\,W/(mK) for HfNiSn thin films compared to literature data on bulk material that varies between $(6.7-10)$\,W/(mK) \cite{Uher99, Hohl99}. As the mismatch between the MgO substrate and HfNiSn is considerably larger than for TiNiSn, the former show wider rocking curves, i.e.\ worse crystallite alignment leading to a reduced thermal conductance. The data shown in this article, however, has been gained on samples with comparable crystalline quality.

Black data points presented in Fig.~\ref{fig:3w}a indicate that indeed a systematic reduction of the thermal conductivity was achieved. The wide range of studied SL periods may be divided into three regions. For large periods, labeled as region 3, we observe $\kappa_{\mathrm{tot}}$ very close to a level marked with a red line that represents the $\kappa_{\mathrm{series}}$ of a bilayer in which the effect of internal interfaces is ignored and the value is calculated from the arithmetic mean of the measured series heat resistances of  the constituent layers 
\begin{equation}
	\kappa_{\mathrm{series}}=\frac{d_1+d_2}{\frac{d_1}{\kappa_{\mathrm{TiNiSn}}+}\frac{d_2}{\kappa_{\mathrm{HfNiSn}}}}
	\label{Eq:KappaSeries}
\end{equation}
where $d_1$ and $d_2$ are the total thicknesses of TiNiSn and HfNiSn.
 One can also notice an apparent saturation of the thermal conductivity for high period lengths, because only few interfaces (15 and 7 for periods equal to 108 nm and 216 nm, respectively) are not significantly affecting $\kappa_{\mathrm{tot}}$. In region 2 the reduction of the thermal conductivity is much more significant and $\kappa_{\mathrm{tot}}$ decreases systematically when the number of interfaces increases from 23 to 575. In this region, an appropriate model is the series of thermal resistances that include both the measured thermal resistivities of the individual films and the interface thermal resistance. 
To simulate the thermal conductivity we apply semiclassical Boltzmann transport, based on the \textit{ab initio} calculation of the bulk phonons using density-functional
perturbation theory \cite{Baroni01}. To get the thermal conductivity one has to integrate the phonon spectra over the total $q$-space summing up all phonon modes $j$. As an approximation we take the phonons as isotropic so that the angular integration gives just a constant.
\begin{equation}
	\kappa_{\mathrm{bulk},x}=\frac{1}{8\pi^2}\sum_j \int_{q_{j,x}}\!\!\!\!\!\!\hbar\omega_{j,x} q^2_{j,x}\left| v(q_{j,x}) \right|^2 \tau_x \frac{dN_0}{dT}dq_{j,x}
\end{equation}
Where $v(q_{j,x})$ is the group velocity of the phonon mode of the material $x$ and $N_0$ the Bose distribution function. It is motivated \cite{Klemens51} that $\tau$ has the form of
\begin{equation}
	\tau_x=\frac{A_x}{\omega^2_{j,x}T}.
\end{equation}
We choose $A_x$ such that the above integral reproduces the experimental bulk thermal conductivity. For the interfaces the mean free path $\ell=v\tau$ must be modified by a transmission coefficient $\zeta$ and the material layer thickness $d$. This is obtained by a diffusive mismatch model \cite{Beechem10}, which is based on the mismatch of the bulk phonons of the two materials. The frequency dependent transmission through the layer and the interface is then given by
\begin{equation}
	\kappa_{x}=\frac{1}{8\pi^2}\sum_j \int_{q_{j,x}}\!\!\!\!\!\!\hbar\omega_{j,x} q^2_{j,x}\left| v(q_{j,x}) \right| \ell(q_{j,x}) \frac{dN_0}{dT} dq_{j,x}
	\label{Eq:vlIntegral}
\end{equation}
with an interface and thickness dependent mean free path
	\begin{equation}
	 \frac{1}{\ell(q_{j,x})} = \frac{1}{d_x\zeta(q_{j,x})}+\frac{\omega^2_{j,x}T}{A_x v(q_{j,x})} 
\end{equation}
Eq.~(\ref{Eq:vlIntegral}) gives the thermal conductivity of material $x$  with layer thickness $d_x$ including the interface through which the phonon leaves the layer and is used in Eq.~(\ref{Eq:KappaSeries}). The green circles in Fig.~\ref{fig:3w}a  that have been obtained from such modeling indicate that the interpretation of the experimental data using thermal interface resistances works for SL periods down to 3 nm (region 2 in Fig.~\ref{fig:3w}a). For thinner SLs the model is no longer valid because the long-wavelength phonons experience an effective medium formed by both materials (see below), which results in a modified phonon dispersion as compared to each individual material.

Finally, for a SL period of 3.2 nm, the lowest thermal conductivity of $(1.11\pm0.06)\,$W/(m\,K) was measured. Decreasing the period even more (region 1), we observe an increase of $\kappa_{\mathrm{tot}}$ due to the formation of an artificial crystal. Along the [100] direction the crystallographic unit cell of a HH compound is composed of $2\times2=4$ atomic layers, where 2 neighboring atomic layers form the chemical unit cell. For the sample with the lowest period of 1.7\,\,nm the new crystal consists of 1.25 unit cells of \hbox{TiNiSn} and 1.5 unit cells of \hbox{HfNiSn}, i.e.\ in total only 11 atomic layers. Locally the chemistry of the HH material requires the crystal size to be 10 or 12 atomic layers, whereas x-rays give the average value. In such a system, phonons experience the material as if it was composed of enlarged unit cells. In this case the still reduced thermal conductance must be attributed to a backfolding of the phonon dispersion relation due to the supercell that introduces gaps in the phonon dispersion relation at the new Brillouin zone boundaries and leads to an overall reduced dispersion of the phonon branches \cite{Tamura99}.

The decomposition of the thermal conductivity into electronic and phononic contribution is summarized in Fig.~\ref{fig:3w}b. The electronic part of the thermal conductivity $\kappa_{\mathrm{el}}$, calculated according to the Wiedemann-Franz law, stays relatively constant for all SL periods, whereas the phononic part follows the trend appointed by $\kappa_{\mathrm{tot}}$. Thereby, the reduction of the cross-plane thermal conductivity can be definitely assigned to the phonons scattering at the interfaces.

\begin{figure}
 \includegraphics[width=3.375in]{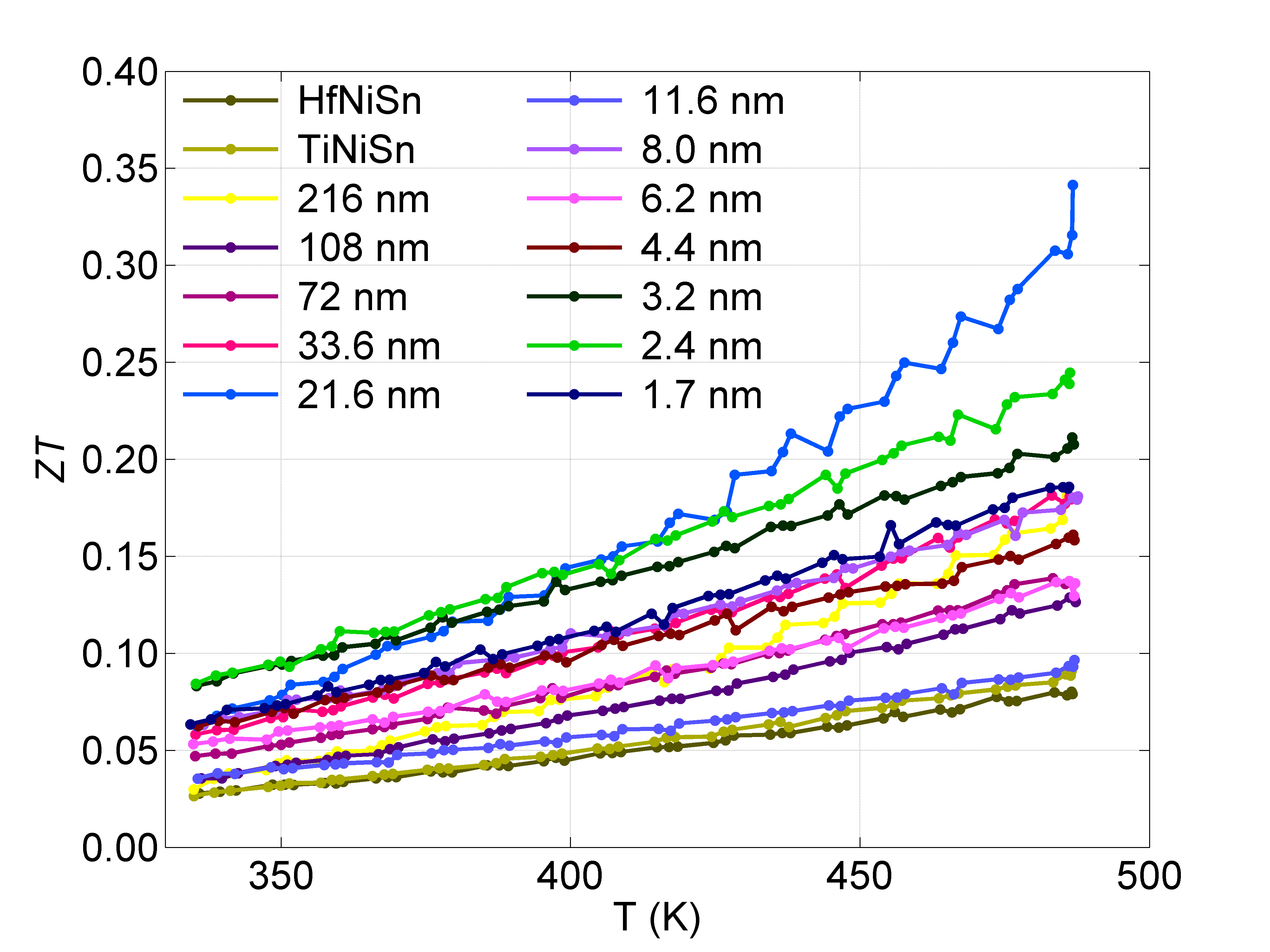}
 \caption{ ZT values calculated from in-plane Seebeck coefficient and electrical resistivity and cross-plane thermal conductivity for TiNiSn, HfNiSn and their SLs.}
 \label{fig:ZT}
\end{figure}

The in-plane resistivity and Seebeck coefficient were measured simultaneously by an LSR 3 (Linseis) in a He atmosphere between room temperature and 480 K. The plot summarizing the measured data is presented in the supplementary material. Due to the isoelectronic substitution we expect a lower interface scattering for electrons than for phonons and assume isotropic electronic transport. This assumption is confirmed by the absence of a systematic correlation between the electronic properties and the interface density. The observed increase of the Seebeck coefficient for very thin layers might be due to some degree of electron confinement in the layers.

Using the room temperature cross-plane thermal conductivity and the temperature dependent in-plane power factor, we estimated the \textit{ZT} values and present them in Fig.~\ref{fig:ZT}. The figure of merit of both single films is almost identical along the entire temperature range and reaches a value of about 0.08 at 480 K, whereas all the studied SLs exhibit greater \textit{ZT} values. In the extreme case, for a SL having a period of 21.6 nm, \textit{ZT} reach a value of almost 0.35, which is 4.4 times larger than the figure of merit of single films. The currently achieved value falls well within the interval demarcated by the previously reported data \cite{Kimura09, Birkel13, Kurosaki05, Downie13, Birkel12, Muta09, Poon11, Galazka14, Schwall13, Geng14} at 480 K.

We deduce that both approaches, top-down (doping and phase separation) as well as bottom-up (SLs) are complementary ways to enhance the figure of merit. While the former has no control over the inclusions size, the latter can vary the thickness of layers with great accuracy. As presented above, the figure of merit of 0.35 for the SL is comparable to the best \textit{ZT} values for bulk half-Heusler alloys reported in the literature at 480 K.

In conclusion, the present research was designed to study the effect of interfaces on the thermal conductivity in HH SLs. DC sputtered TiNiSn/HfNiSn SLs reveal a smooth surface and well defined layered structure as proved by AFM, XRD and STEM. The thermal conductivity of thin films was measured using the differential $3\omega$ method. Decreasing SL period, i.e.\ increasing number of interfaces, the expected reduction of the thermal conductivity is observed. For a SL period of 3.2 nm the minimum $\kappa_{\mathrm{tot}}=(1.11\pm0.06)\,$W/(mK) was achieved. Further decrease of the SL period leads to the rise of $\kappa_{\mathrm{tot}}$ due to the creation of an artificial crystal. We calculated the thermal conductivity using semiclassical Boltzmann transport, based on the \textit{ab initio} calculation of the bulk phonons using density-functional perturbation theory and a diffusive mismatch model for phonon scattering at the interfaces. Experimental data and calculation agree for SL periods larger than 3 nm. An electron confinement might be the reason for the improved Seebeck coefficient in SLs with very low periods. The figure of merit was estimated based on the room temperature cross-plane thermal conductivity and the temperature dependent in-plane power factor. Due to the reduced thermal conductivity, the superlattices with all studied periods reveal enhanced \textit{ZT} compared to single constituent films.
\\

We gratefully acknowledge financial support by DFG Ja821/4-2 SPP 1386, BA4171/2-2 SPP 1386, DFG SPP 1538 'Spin Caloric Transport' and the Graduate School of Excellence Material Science in Mainz (GSC 266).

\bibliography{TNS_HNS_paper}

\end{document}



\title{Reduced thermal conductivity of TiNiSn/HfNiSn superlattices - \\supplementary data}

\author{Paulina Ho\l{}uj}
 \email{holuj@uni-mainz.de}
 \affiliation{Institute of Physics, University of Mainz, Staudinger Weg 7, 55128 Mainz, Germany.}
 \affiliation{Graduate School Materials Science in Mainz, Staudinger Weg 9, 55128 Mainz, Germany.}

\author{Christoph Euler}
 \affiliation{Institute of Physics, University of Mainz, Staudinger Weg 7, 55128 Mainz, Germany.}

\author{Benjamin Balke}
 \affiliation{Institute of Inorganic and Analytical Chemistry, University of Mainz, Duesbergweg 10-14, 55128 Mainz, Germany.}

\author{Ute Kolb}
 \affiliation{Institute for Physical Chemistry, University of Mainz, Welderweg 11, 55099 Mainz, Germany.}

\author{Gregor Fiedler}
 \affiliation{Faculty of Physics, University of Duisburg-Essen, Lotharstra{\ss}e 1, 47048 Duisburg, Germany.}

\author{Mathis M. M\"uller}
 \affiliation{Institute of Applied Geosciences, TU Darmstadt, Schnittspahnstra{\ss}e 9, 64287 Darmstadt, Germany.}

\author{Tino Jaeger}
 \affiliation{Institute of Physics, University of Mainz, Staudinger Weg 7, 55128 Mainz, Germany.}

\author{Peter Kratzer}
 \affiliation{Faculty of Physics, University of Duisburg-Essen, Lotharstra{\ss}e 1, 47048 Duisburg, Germany.}

\author{Gerhard Jakob}
 \affiliation{Institute of Physics, University of Mainz, Staudinger Weg 7, 55128 Mainz, Germany.}
 \affiliation{Graduate School Materials Science in Mainz, Staudinger Weg 9, 55128 Mainz, Germany.}

\date{\today}

\maketitle

\section{Sample preparation}
TiNiSn/HfNiSn superlattices (SLs) were prepared by DC magnetron sputtering at 520$^{\circ}$C and an Ar pressure of 3.1 Pa on previously annealed MgO (001) substrates. While the cathodes were operating continuously at 23 W (TiNiSn) and 20 W (HfNiSn), a heater with the attached sample holder was moving back and forth between both cathodes to create a layered structure. The movement of the heater was optimized in a way that compensates a possible thickness gradient caused by multiple approaching of the cathode from the same side. Based on the determined deposition rate the waiting time over the respective target was calculated so that the thickness of both layers in one period was the same.

\section{3$\omega$ method}
The cross-plane thermal conductivity of half-Heusler (HH) SLs was measured with the 3$\omega$ method \cite{Cahill90, Cahill02, Borca-Tasciuc01}. On top of each SL, 200 nm of Al$_{2}$O$_{x}$ was deposited by RF magnetron sputtering to electrically insulate the film from the heater/thermometer line, which was produced in an optical lithography step from a 50 nm thick Au layer. Such a 20 $\mu$m wide heater line was terminated with two electrodes that are used to apply an AC current. Two additional pads were connected to the heater in between. An alternating current with frequency $\omega$, applied to the outer electrodes, passes through the metal strip causing periodic Joule heating and modulating its temperature dependent resistance with a frequency of 2$\omega$. Consequently, according to Ohm's law, the voltage drop between the inner connections contains contributions proportional to the first and third harmonic. The usage of the differential variant of the 3$\omega$ method \cite{Borca-Tasciuc01} requires preparation of two samples - regular (1000\,\,nm thick) and reference (150\,\,nm thick) for every SL period. Assuming the one dimensional heat conduction model, the thermal conductivity of the film is expressed by Eq. (\ref{eq:eq1}):
	\begin{equation} \label{eq:eq1}
	\kappa_{\mathrm{film}}=\frac{P\cdot(d_{\mathrm{regular}}-d_{\mathrm{reference}})}{2b\cdot l\cdot(\Delta T_{\mathrm{regular}}-\Delta T_{\mathrm{reference}})}
	 \end{equation}
where $P$ is the power applied to the heater, $d$ is the thickness of the respective layers, $2b$ is the width of the heater line, $l$ is the distance between the voltage probes and $\Delta$$T$ is represented by Eq. (\ref{eq:eq2}). It contains the third harmonic of the voltage drop $U_{3\omega}$, the voltage across the heater line at frequency $\omega$, i.e. $U_{\omega}$, the resistance of the heater $R$ and and its temperature derivative.
	\begin{equation} \label{eq:eq2}
	\Delta T=2\frac{U_{3\omega}}{U_{\omega}}\frac{R}{\frac{dR}{dT}}
	 \end{equation}

\section{The thermoelectic quantities}
\begin{figure*}
 \includegraphics[width=5.75in]{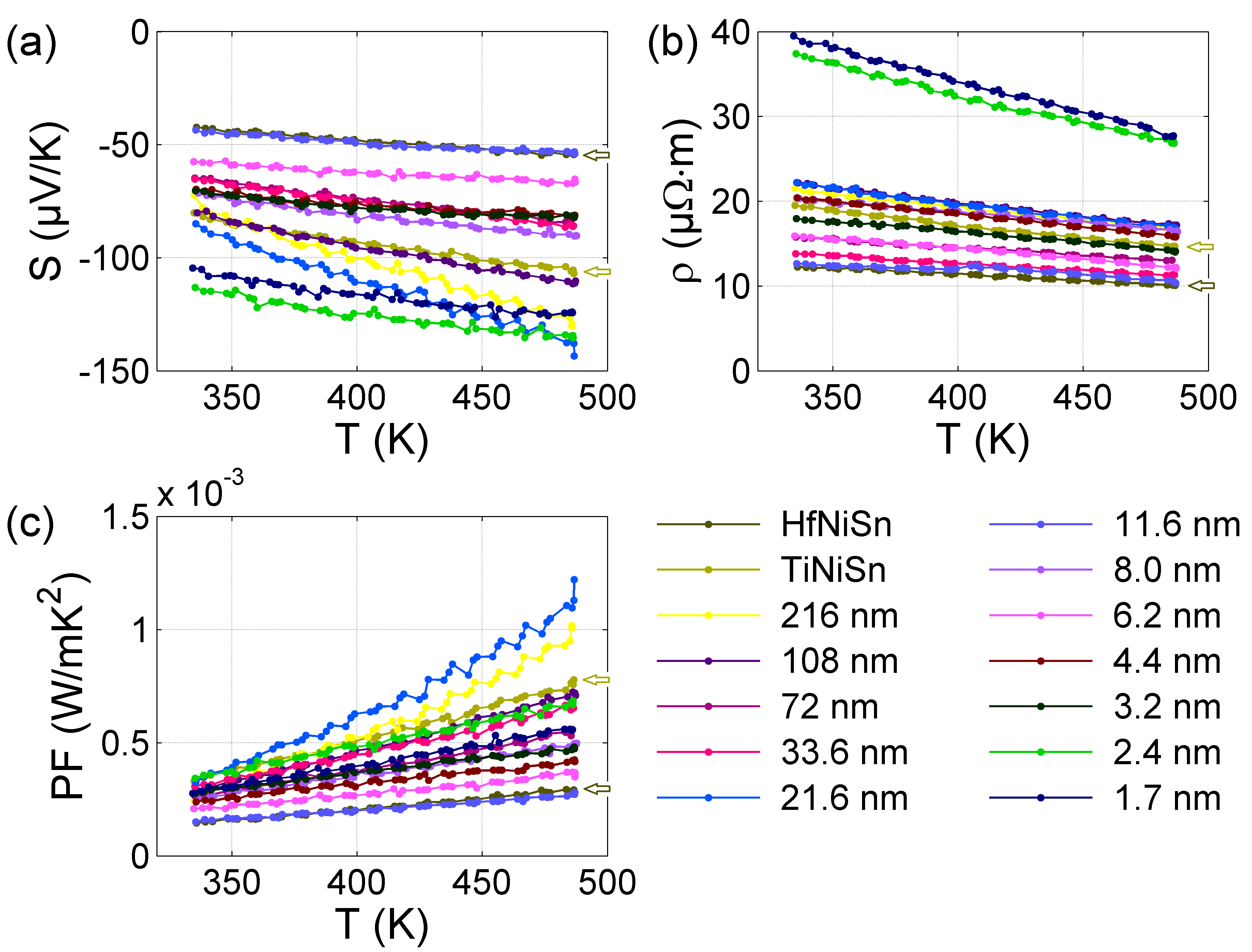}
 \caption{ (a) In-plane Seebeck coefficient, (b) in-plane electrical resistivity and (c) power factor for the SLs with a given period and single materials TiNiSn and HfNiSn.}
 \label{fig:Seebeck}
\end{figure*}

The in-plane resistivity and the Seebeck coefficient were measured simultaneously by an LSR 3 (Linseis) in a He atmosphere between room temperature and 480 K. Both of these quantities and resulting power factor are summarized in Fig.~\ref{fig:Seebeck}. The horizontal arrows indicate the data measured for TiNiSn and HfNiSn single films keeping the color code presented in the legend. Due to the isoelectronic substitution on the \textit{M} site of \textit{M}NiSn we expect a lower interface scattering for electrons than for phonons and assume isotropic electronic transport. This assumption is confirmed by the absence of a systematic correlation between the electronic properties and the interface density.

However, an interesting finding is an improved Seebeck coefficient for SLs having extremely low periods. Although we used isoelectronic materials, due to the different lattice constants a residual Fermi surface mismatch exists that can lead to some degree of electron confinement and an enhanced Seebeck coefficient.

Having a closer look on the power factor (Fig.~\ref{fig:Seebeck}c) one can notice that the initial assumption about the fact that electronic parameters $S^2\sigma$ should not be affected by the SL approach is valid. None of the investigated SL samples exhibited lower power factor than the single HfNiSn film.

\bibliography{TNS_HNS_paper}